\documentstyle[epsf]{article}
\makeatletter
\def\epsfile#1{\def\@psfile{}\parse@ps@parms{#1}}
\def\parse@ps@parms#1{%
  \@for\@epsfile:=#1\do{\expandafter\@setparms\@epsfile,}%
  \epsfbox{\@psfile}}
\def\@setparms#1=#2,{\@nameuse{@setps#1}{#2}}
\def\@setpsfile#1{\def\@psfile{#1}}
\def\@setpsheight#1{\epsfxsize=0pt\epsfysize=#1}
\def\@setpswidth#1{\epsfxsize=#1\epsfysize=0pt}
\def\@setpsscale#1{\def\epsfsize##1##2{#1##1}}
\makeatother
\begin{document}
\begin{centering}
{\Large \bf Analysis of low pressure electro-positive and
electro-negative rf plasmas with
Monte Carlo method}
\\ \vspace{.75in}
\Large {M. Ardehali}
\\  \vspace{.5in}
\large {Silicon Systems Research Laboratories,
NEC Corporation,
Sagamihara,
Kanagawa 229
Japan}
\\ \vspace{.75in}
\end{centering}

\begin{abstract}
Particle-in-cell/Monte Carlo technique is used to simulate low pressure
electro-negative and electro-positive plasmas
at a frequency of 10 MHz. The potential, electric field,
electron and ion density, and
currents flowing across 
the plasma are presented.
To compare the physical properties of the electro-positive gas with
those of an electro-negative gas, the input voltage was
decreased from $1000$ Volts to $350$ Volts.
The simulation results indicate that the
introduction of negative ions induces dramatic effects on the 
spatial and temporal variation of the
electric field and on the
behavior of the electron, ion, and displacement currents.
In particular, the numerical modeling predicts the formation of 
double-layer at the plasma-sheath boundary of electro-negative 
discharge.
\end{abstract}

\pagebreak

\begin{center}
\Large  {\bf I. Introduction}
\end{center}

Plasma-assisted processing
is becoming increasingly important in microelectronics industry
\cite {1}. The
figures-of-merit of the process (for example
anisotropy, etching rate, etc.)
are strongly influenced by the discharge properties, such as
plasma density,
plasma potential, etc \cite {2}, \cite {3}. 
Obviously predictive
models are needed to analyze the rf glow discharges \cite{4}.
Unfortunately, plasma discharges
are complex systems 
that are very difficult to analyze. In recent years,
there has been considerable efforts
to model these systems self-consistently.
Particle-in-cell/Monte-Carlo (PIC/MC) simulation
has proven to be an effective and
powerful tool in increasing the general understanding of plasma
processing.  PIC/MC simulation is very attractive because it has 
the advantage of being
self-consistent and does not require any
assumptions regarding the
internal electric field, the electron velocity distribution function,
and the relaxation time approximation
\cite{5}.

The purpose of the present paper
is to compare the physical properties of 
an electro-positive gas with those of 
an electro-negative gas at comparable 
plasma densities. The simulation results indicate 
that the introduction of negative 
ions in rf plasma induces dramatic effects on the electric field
distribution within the discharge and leads 
to the double-layer formation.
\begin{center}
\Large  {\bf II. Details of simulation}
\end{center}

We have developed a PIC/MC simulator to model electro-positive and
electro-negative discharges. Monte Carlo technique is
considerably more accurate than fluid 
dynamics method because it accounts for the
non-local effects naturally; whereas fluid dynamics
accounts for the non-local effects on an {\em ad hoc} basis.

Very briefly,
the present PIC/MC simulator
consists of the following steps:
\\
(1) The instantaneous locations of MC particles
representing ions
and electrons are
interpolated to the grid points to obtain the charge density.
\\
(2) Poisson's equation on a
spatially discretized mesh is solved to obtain the electric field.
\\
(3) The electric field is interpolated from the grid points to the
location of MC particles.
\\
(4) The equations of motion under the local and
instantaneous electric field are integrated.
\\
(5) Random numbers (Monte Carlos technique)
and collision cross sections are used to
determine the probability that each particle suffers
a collision.

As an example of the application
of this simulator, we have modeled an rf plasma at
a pressure of $20$ mTorr.
Since in this work we are primarily interested in comparing the physical
properties of the electro-positive and electro-negative gases,
we have intentionally
kept the plasma model simple.
In the simulation, the applied frequency
is $\frac{\displaystyle\omega}{\displaystyle2\pi}=10$
MHz, 
 and the discharge length is $20$ cm with perfectly
absorbing electrodes.
The time
step $\delta t=0.1$ ns which is small enough to resolve the electron
plasma frequency.
For the electro-positive plasma,
an Argon type discharge was considered.
The left electrode of the discharge is driven
with a voltage $V_{rf}(t)= V_{rf} sin\omega t$, where
$V_{rf}= 1000$ Volts.
The electron-neutral
ionization cross section $\sigma(v)$ is given by
$\sigma(v)=\frac{\displaystyle k}{\displaystyle v}$,
where $v$ is the electron
velocity and the rate constant $k$
is $2 \times 10^{-8}$ $cm^3/s$.
Ionizing collisions occur if the electron energy is larger than
$15$ eV.
An ionizing collision is modeled by loading a new electron
and ion at the position of the ionizing electron. The kinetic
energy after ionizing  collision is partitioned between the two
electrons with equal probability.

For the electro-negative discharge,
a Chlorine gas was considered.
The left electrode of the discharge is driven
with a voltage $V_{rf}(t)= V_{rf} sin\omega t$, where
$V_{rf}= 350$ Volts.
The ionization cross section is the same as for the electro-positive
gas.
The additional
processes considered are dissociative electron attachment
leading to negative ion formation
and positive ion/negative ion recombination.
The scattering and ionization processes are represented as follows:
\begin{eqnarray}
Cl_2 + e \rightarrow Cl^- + Cl\\ \nonumber
Cl_2 + e \rightarrow Cl_{2}^{+} + 2e \\ \nonumber
Cl_{2}^{+} + Cl^{-} \rightarrow Cl_{2} + Cl
\end{eqnarray}
where $Cl_{2}^{+}$ and $Cl^{-}$ are the main positive and
negative ions, respectively. The rate constants for attachment
and recombination are $1.8 \times 10^{-10}$ $cm^3/s$ and
$5 \times 10^{-8}$ $cm^3/s$ respectively
\cite {6}.
In the electro-negative discharge,
the electron density
is an order of magnitude smaller
than the positive (or negative) ion density;
in the simulation, however,
the electron weight was adjusted to
make the number of
MC particles corresponding to electrons similar to the number
of MC particles corresponding to positive and negative ions.

\begin{center}
\Large  {\bf III. Electro-positive discharge}
\end{center}

We first present the simulation results for
an electro-positive discharge at a pressure of 
20 mTorr.
Figures $1 (a)$ and $1 (b)$ show
the potential and the electric field
at four times during the rf cycle.
In the bulk, the
potential is constant, and hence the
electric field is negligible. In the
sheath, however, the
potential changes rapidly
(due to accumulation of positive charge in the space charge region),
and the electric field therefore
becomes very large, especially near
the electrodes. 
This is particularly evident at 
$\frac{\displaystyle \omega t}{\displaystyle 2 \pi}=0.25$ and
$\frac{\displaystyle \omega t}{\displaystyle 2 \pi}=0.75$.
Figures $2 (a)$ and $2 (b)$ show the ion and electron
density of the electro-positive
plasma.  The ion density is independent of the phase of the
applied voltage, whereas the electron density oscillates within the
sheath. The large electric field at the sheath pushes
the electrons away from the sheath, whereas it 
accelerates some of the positive ions
toward the electrodes.

Figures $3 (a)$,
$3 (b)$, $3 (c)$, and $3 (d)$ show the spatial and temporal
variations of the
electron current, 
displacement current, 
ion current, 
and total current
at four times during the cycle respectively
(the units are not relative and represent true values).
At 
$\frac{\displaystyle \omega t}{\displaystyle 2 \pi}=0$ and
$\frac{\displaystyle \omega t}{\displaystyle 2 \pi}=0.5$,
the sheath current is predominantly
displacement current, whereas at
$\frac{\displaystyle \omega t}{\displaystyle 2 \pi}=0.25$ and
($\frac{\displaystyle \omega t}{\displaystyle 2 \pi}=0.75$)
the electron
current is the
dominant current in the sheath of the powered (grounded)
electrode.
The electron current, however, is dominant in the
bulk at all times during the cycle.
Since
the bulk electric field does not change with time,
the displacement current has a
negligible value in the bulk.
Also note that the bulk ion current (Fig. 3 (c)) does not modulate
with time, but
at the sheath boundary,
the ion current 
increases abruptly 
due to the large sheath
electric field and oscillates with time. The total current, which
is defined as the sum of the electron, 
displacement and ion current (Fig. 3 (d)),
is constant across the discharge.

\begin{center}
\Large  {\bf IV. Electro-negative discharge}
\end{center}

To obtain an electro-negative discharge with plasma density
comparable to that
of the electro-positive discharge, the applied voltage was
decreased from 1000 Volts to 350 Volts.
Figure $4 (a)$ shows the potential at four times during the rf
cycle. The 
potential changes significantly near the electrodes, leading to large
electric fields within the sheaths. However unlike the electro-positive
plasma, the potential in the bulk is not constant.
Figure $4 (b)$ shows the electric field
profile at four times during the cycle.
Since negative
ions and electrons are excluded from the sheath,
the space charge within the sheath 
is positive. Note that at times
$\frac{\displaystyle \omega t}{\displaystyle 2 \pi}=0.25$
and
$\frac{\displaystyle \omega t}{\displaystyle 2 \pi}=0.75$
the electric field
becomes very large near
the electrodes.

Comparing Figs. 1 (b) and 4 (b), 
three important differences between the
electro-positive and
electro-negative discharges may be noted:
\\
(1) Unlike the electro-positive discharge, 
the electric field has a nonvanishing value
in the bulk of the electro-negative discharge.
\\
(2) In electro-negative discharge,
the electric field shows a relative maximum at the plasma-sheath
boundary. The maximum is due to double-layer formations and is
particularly important at
$\frac{\displaystyle \omega t}{\displaystyle 2 \pi}=0$ and
$\frac{\displaystyle \omega t}{\displaystyle 2 \pi}=0.5$. This 
result is in agreement
with experimental measurements \cite{7}.
\\
(3) The magnitude of the sheath electric field
is significantly larger
in the electro-positive discharge than in the 
electro-negative plasma.
\\
These factors have profound influence on the behavior of the
spatial and
temporal variations of the electron, ion, 
and displacement currents within
the discharge.

Figures $5 (a)$ and $5 (b)$ show 
the positive and negative ion density within the 
discharge. The ion density is independent of the phase of the 
applied voltage. The large electric field at the sheath pushes
the negative ions away from the sheath and hence negative ions are 
completely excluded from the sheath. However, some of the positive ions
are accelerated by the sheath toward the electrodes.
It should be noted that in the
bulk, negative ions are lost due to recombination and hence
do not accumulate at the boundary of the sheath and 
the glow region. Figure $5 (c)$ shows the electron density at four times
during the cycle. Unlike the ions, the electron density is a function 
of time and varies significantly within the sheath. When the sheath 
electric field at the left electrode is small, 
electrons move toward the left electrode but are excluded from the
right electrode. On the other hand when the sheath
electric field at the right electrode is small,
electrons move toward the right electrode but
are excluded from the left electrode. The modulation of the
bulk electron density, however, is small and almost negligible.
Fig. $5 (d)$ compares the electron, positive ion and negative ion a
density at four times during the cycle. Note that in the bulk,
the sum of the
electron density and negative ion density is approximately equal to the
positive ion density and hence
total charge 
within the bulk is almost zero.

Figures $6 (a)$, $6 (b)$, $6 (c)$, $6 (d)$, and $6 (e)$ 
show the spatial and temporal
variations of
the electron current, displacement current,
positive ion current,  negative ion current,
and total current
at four times during the cycle respectively
(again, the units are not relative and represent true values).
Similar to the electro-positive discharge, at
$\frac{\displaystyle \omega t}{\displaystyle 2 \pi}=0$ and
$\frac{\displaystyle \omega t}{\displaystyle 2 \pi}=0.5$
the sheath current is predominantly
displacement current, whereas at
$\frac{\displaystyle \omega t}{\displaystyle 2 \pi}=0.25$
($\frac{\displaystyle \omega t}{\displaystyle 2 \pi}=0.75$)
the electron
current is the dominant current in the sheath of the powered (grounded)
electrode. The
electron current is dominant in the
bulk at all times during the cycle.
Note that the displacement current within the bulk
is not zero and contributes to the total current. The non-vanishing of
the bulk displacement current is
primarily due to temporal 
variation of bulk electric field [see Fig. $4 (b)$;
the displacement current $J_d(t)$ is defined as
$J_d(t)=-\epsilon_0(dE/dt)$].
The positive ion current across the discharge is shown in Fig. $6 (c)$.
The positive ion current
increases only slightly at the sheath boundary.
Comparing Figs. 3 (c) and 6 (c), one can see that in the 
electro-negative discharge, the ion current in the bulk
varies with time (this
is primarily due to temporal variation of the bulk electric field),
and has a smaller modulation within the sheath.
Fig. $6 (d)$ shows 
the negative ion current across the discharge.
Since negative ions
are confined to the bulk of the plasma,
the current  within the sheath is
negligible.
However, in the bulk, the negative ion current
oscillates with the applied voltage.
Note the small contribution of
positive and negative ion currents to
the total current. Finally
Fig. $6 (e)$ shows the total current,
which is
defined as sum of the
electron, positive  and negative ion, and displacement
current across the discharge.
Although there is a small blip at the sheath
boundary, 
the total current within the discharge is constant
(experience has
shown that the blip diminishes by
increasing computation time).

It is worth noting that
although plasma density in the two cases are comparable,
the total current (as well as the
electron, ion and displacement current) 
is approximately
three times
smaller
in the electro-negative discharge
than the corresponding currents in
the electro-positive plasma.

\pagebreak
\begin{center}
\Large  {\bf V. Conclusion}
\end{center}

In summary, we have presented the PIC/MC simulation results of an
electro-positive and an electro-negative discharge at comparable
plasma densities. The numerical results indicate that
the presence of negative ions leads 
to a relative maximum at the plasma-sheath
boundary, i.e., to the formation of double layers. This result is
in agreement
with the experimental measurements.
The simulation results also indicate that
unlike the electro-positive discharge,
the bulk electric field of the
electro-negative plasma
has a nonvanishing value and oscillates with time. This oscillation
leads to significant
bulk displacement current and to modulation of bulk positive ion 
current.

Simulation results such as the ones reported here
show great promise in interpreting experimental measurements and
in testing the validity of analytical models and
simulation
models based on fluid dynamics. Future experiments, in conjunction with
similar Monte Carlo simulations,
can provide considerable information
about the details of rf discharges and about the
formation of double-layers.

\pagebreak

\pagebreak
\noindent
{\em \bf Important Note:} All figures should be
considered from left to right.
\\
Figures 1 (a), 1 (b), 
2 (a), and 2 (b) are on the same page.
\\
Figures 3 (a), 3 (b), 
3 (c), and 3 (d) are on the same page.
\\
Figures 4 (a)
and 4 (b) are on the same page.
\\
Figures 5 (a),
5 (b), 5 (c), and 5 (d) are on the same page.
\\
Figures 6 (a),
6 (b), 6 (c), 6 (d), and 6 (e) are on the same page.
\\
{\bf Figure $1 (a)$-}
The spatial and temporal variations of the potential in the discharge.
For Figs. 1 (a) to 3 (d), the discharge is electro-positive. 
All figures represent profiles at four times during the rf cycle:
$\frac{\displaystyle \omega t}{\displaystyle 2 \pi}=0$ (solid line),
$\frac{\displaystyle \omega t}{\displaystyle 2 \pi}=0.25$ 
(dotted line),
$\frac{\displaystyle \omega t}{\displaystyle 2 \pi}=0.5$ 
(dashed-dashed line),
$\frac{\displaystyle \omega t}{\displaystyle 2 \pi}=0.75$ (dashed line).
 the electro-negative gas 
\\
{\bf Figure $1 (b)$-}
The electric field
profile.
\\
{\bf Figure $2 (a)$-}
The positive ion density.
\\
{\bf Figure $2 (b)$-}
The electron density.
\\
{\bf Figure $3 (a)$-}
The spatial and temporal variations of the electron current across
the discharge.
\\
{\bf Figure $3 (b) $-}
Same as Fig 3 (a) but for  
the displacement current.
\\
{\bf Figure $3 (c)$-}
Same as Fig 3 (a) but for  
the ion current.
\\
{\bf Figure $3 (d)$-}
The spatial and temporal variations of the total current across the
discharge.
\\
{\bf Figure $ 4 (a)$-}
The potential profile in the discharge.
For Figs. 4 (a) to 6 (e), the discharge is electro-negative.
\\
{\bf Figure $4 (b)$-}
The electric field profile.
\\
{\bf Figure $5 (a)$-}
The positive ion density.
\\
{\bf Figure $5 (b)$-}
The negative ion density.
\\
{\bf Figure $5 (c)$-}
The electron density.
\\
{\bf Figure $5 (d)$-}
The positive ion, negative ion, and
electron density.
\\
{\bf Figure $6 (a)$-}
The spatial and temporal variations of the electron current across
the discharge.
\\
{\bf Figure $6 (b)$-}
Same as Fig 6 (a) but for
the displacement current.
\\
{\bf Figure $6 (c)$-}
Same as Fig 6 (a) but for
the positive ion current.
\\
{\bf Figure $6 (d)$-}
Same as Fig 6 (a) but for
the negative ion current.
\\
{\bf Figure $6 (e)$-}
The spatial and temporal variations of the total current across
the discharge.
\pagebreak

\epsfile{file=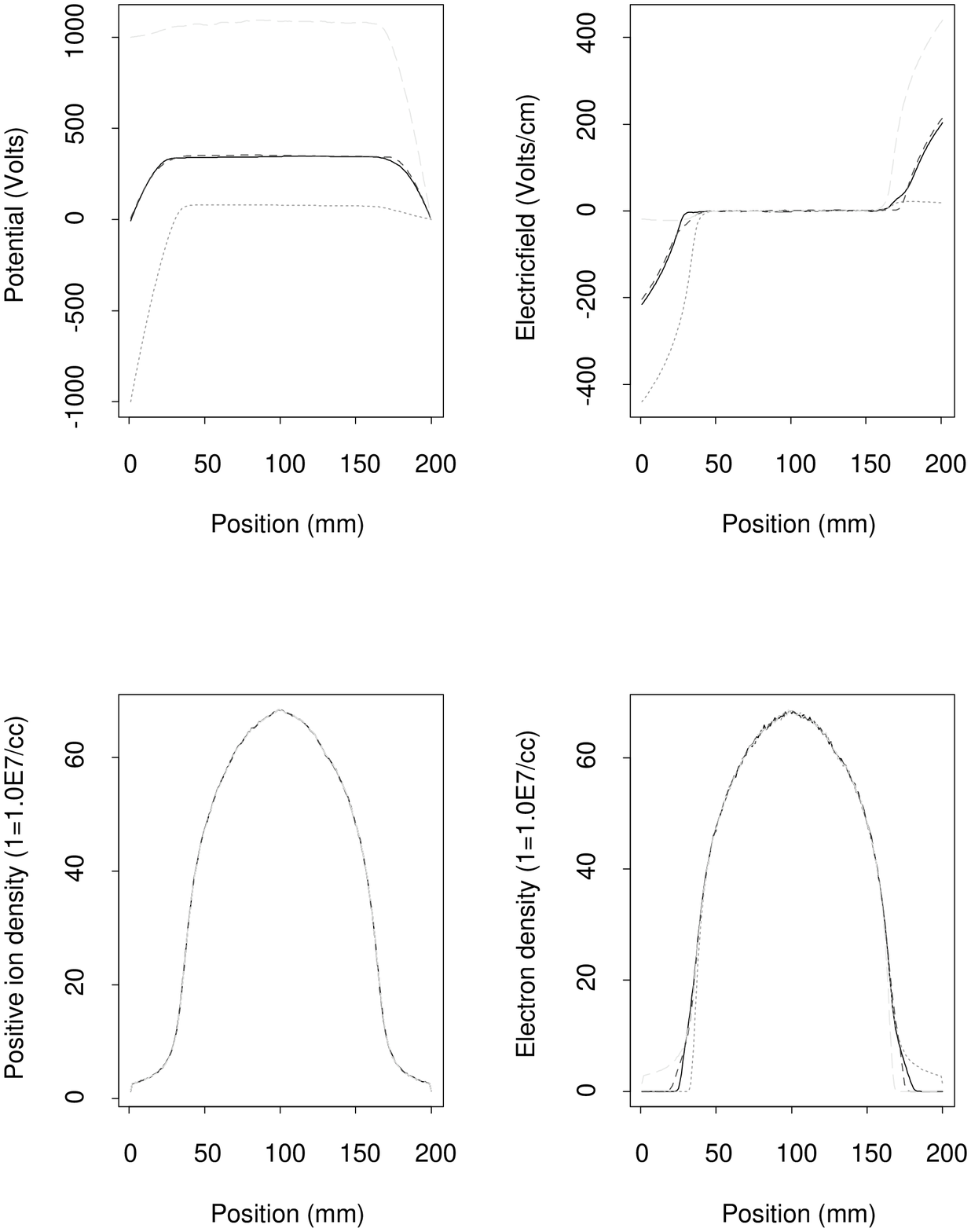,width=\textwidth}
\epsfile{file=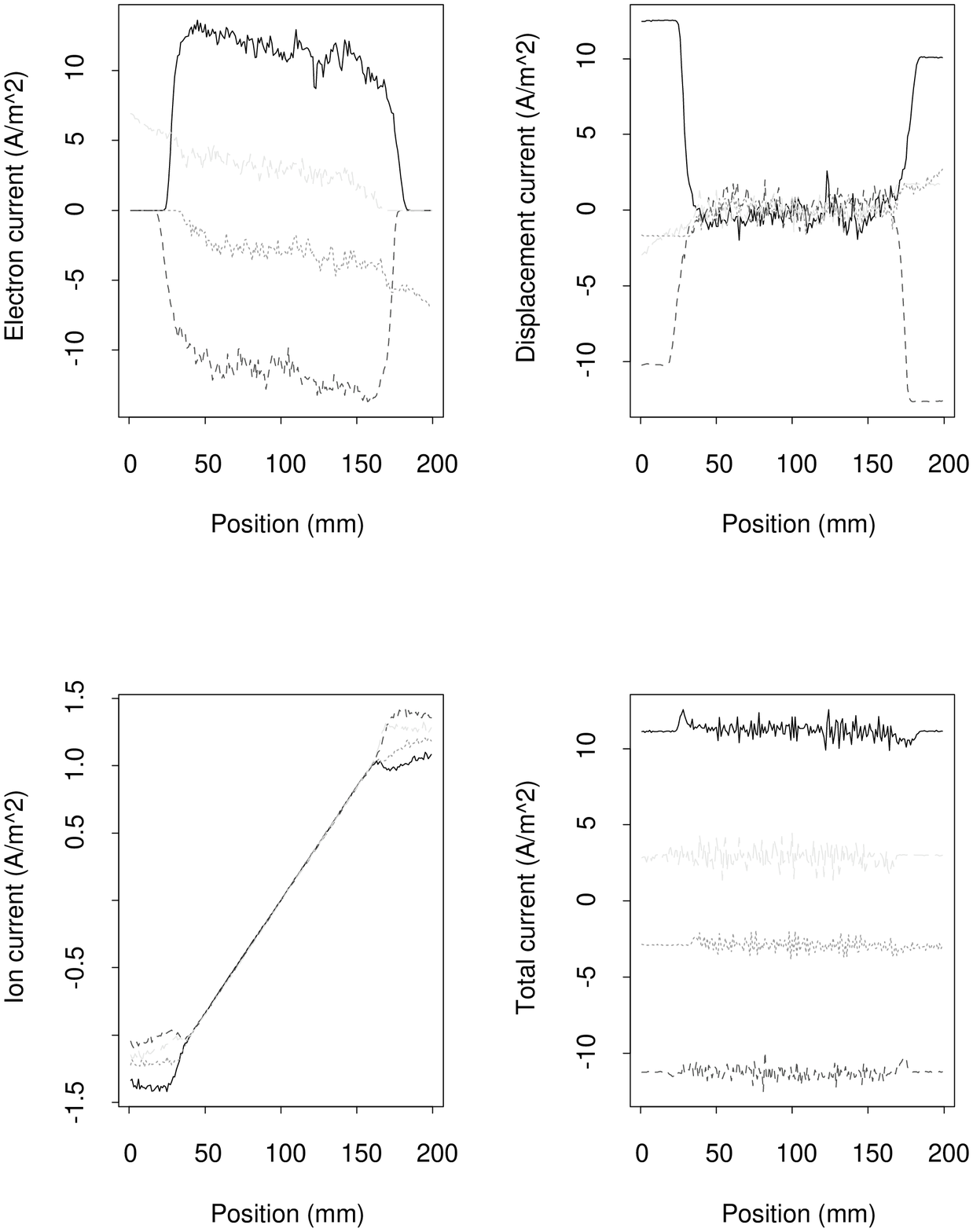,width=\textwidth}
\epsfile{file=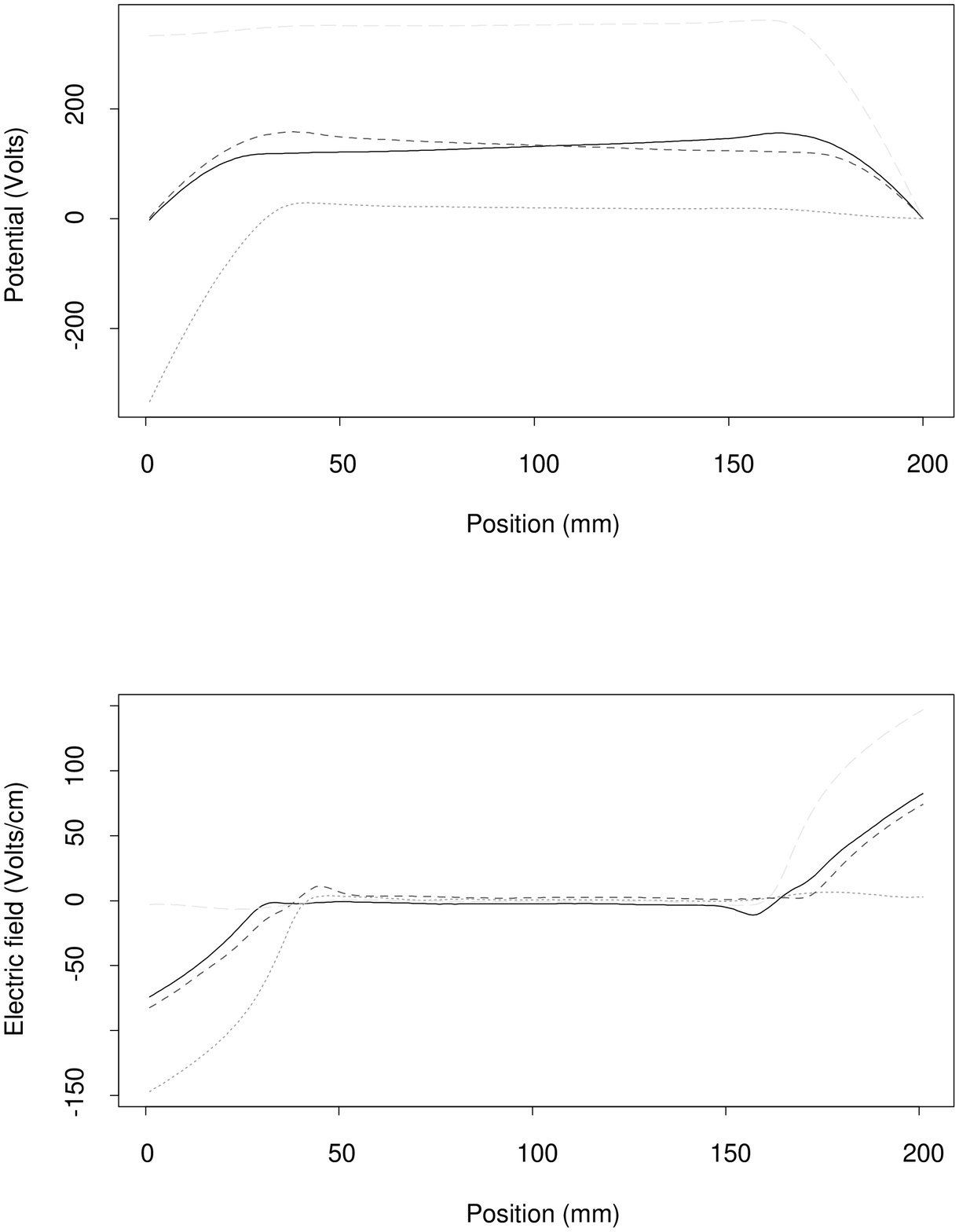,width=\textwidth}
\epsfile{file=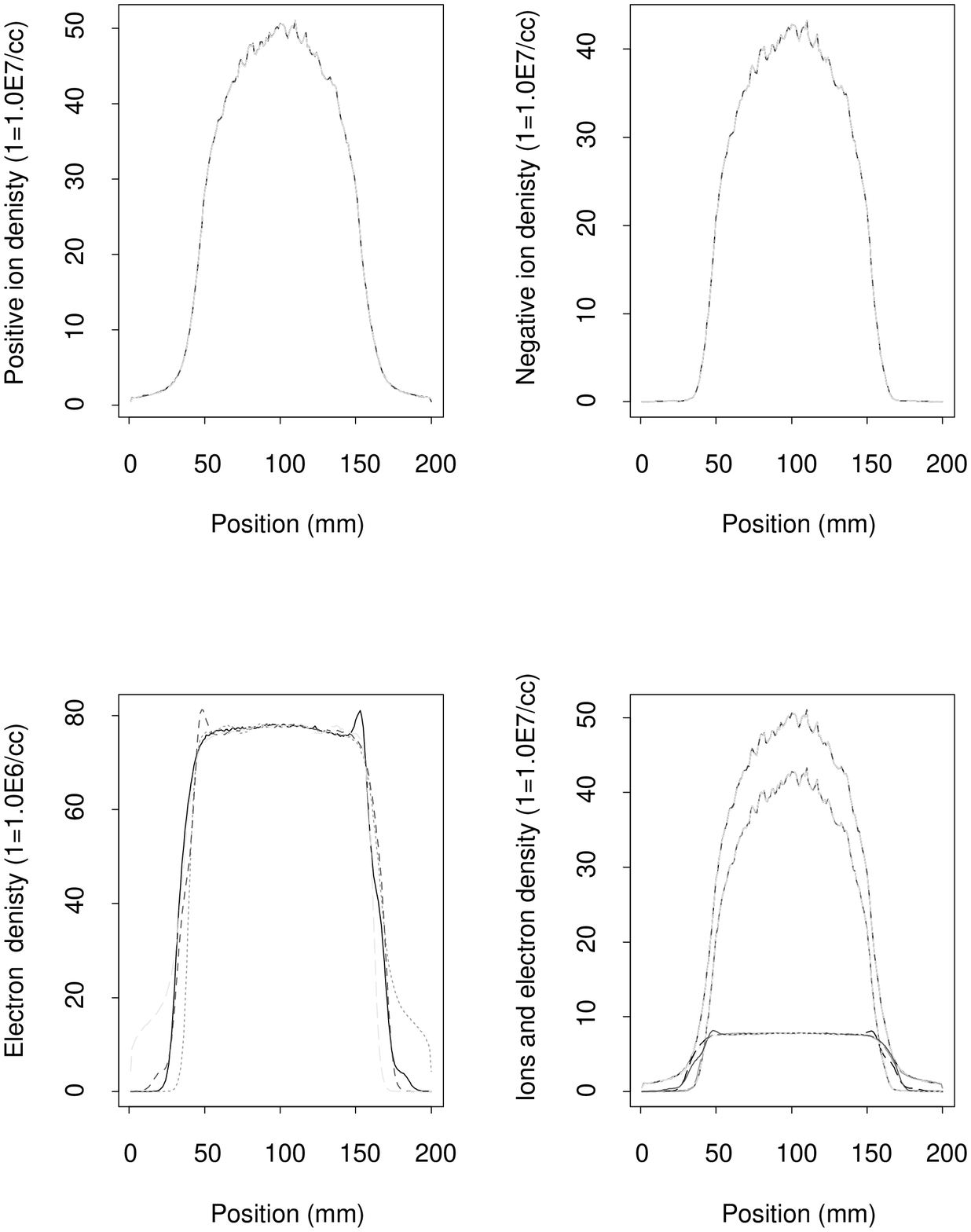,width=\textwidth}
\epsfile{file=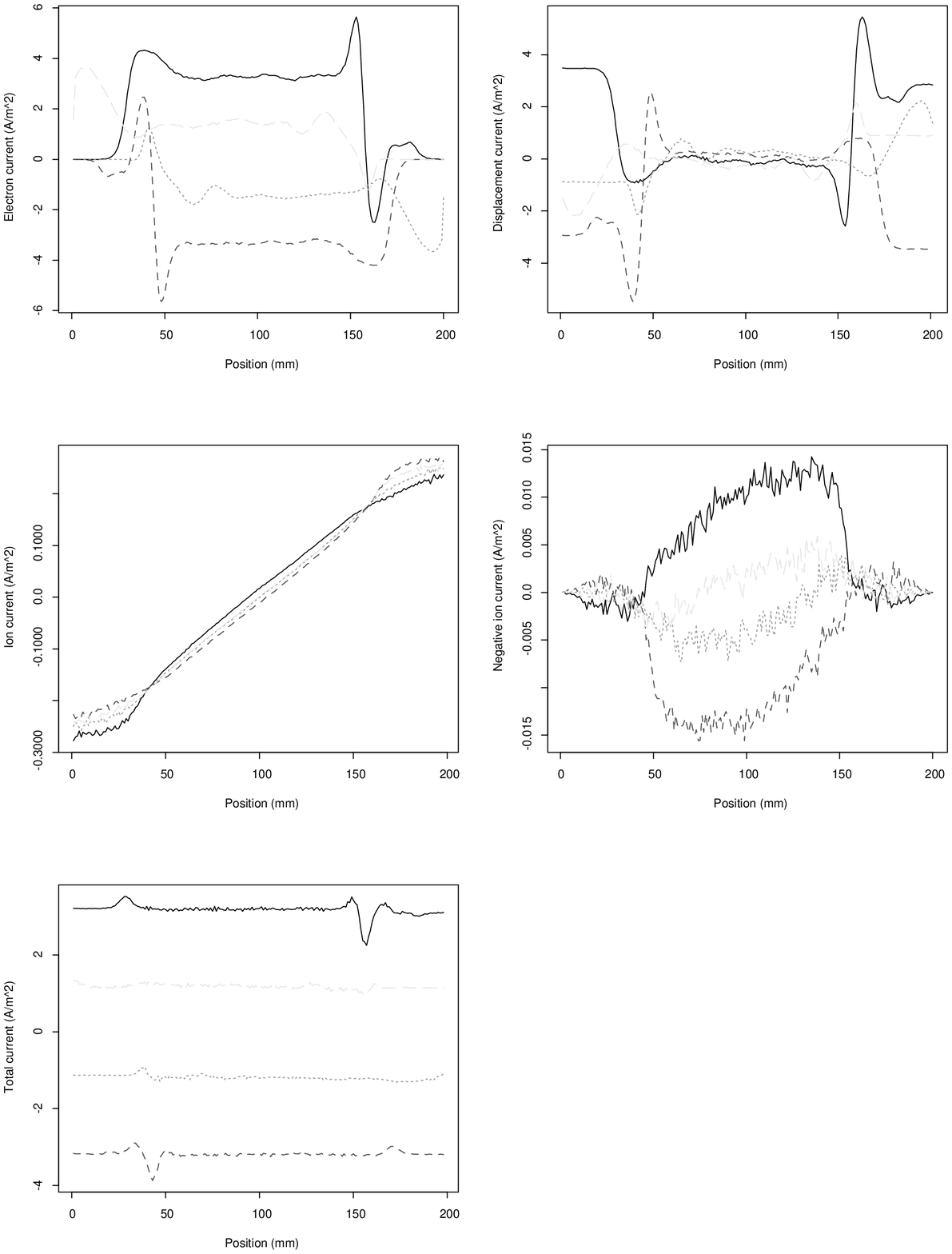,width=\textwidth}


\begin {thebibliography}{99}
\bibitem{1} D. M. Manson and D. L. Flamm, {\it
Plasma Etching} (Academic, New York, 1989) p, 12.

\bibitem{2}
R. A. Gottscho, C. J. Jurgensen, and D. J. Vitkavage,
``Microscopic uniformity in plasma etching,''
{\it J. Vac. Sci. Technol. B}, vol. 10, pp. 2133-2147, 1992.

\bibitem {3} M. Ardehali, ``Monte Carlo simulation of ion transport
through radio frequency collisional sheath,'' 
{\it J. Vac. Sci. Technol. A}
vol. 12, pp. 3242-3245, 1994;
M. Ardehali, ``Effect of image force on ion current density in plasma
discharges'',  {\it IEEE Trans. Plasma
Sci.}, vol. 24, pp. 241-245, 1996.

\bibitem{4} E. Gogolides, J. P, Nicolai, and H. H. Sawin,
``Comparison of experimental measurements and model predictions
for radio-frequency Ar and $SF_6$ discharges,'' {\it
J. Vac. Sci. Technol. A}, vol. A7, no. 3, pp. 1001-1005, 1989;
E. Gogolides, and H. H. Sawin,
"Continuum modeling of radio-frequency glow discharges. I. Theory
and results for electropositive and electronegative gases,"
{\it J. Apl. Phys.}, vol 72, pp. 3971-3987, 1992;
E. Gogolides, and H. H. Sawin,
"Continuum modeling of radio-frequency glow discharges. II.
Parametric studies and sensitivity analysis,"
{\it J. Apl. Phys.}, vol 72, pp. 3988-4002, 1992.

\bibitem{5}
D. Vender and R. W. Boswell, ``Numerical
modeling of low-pressure rf plasmas,'' {\it IEEE Trans. Plasma
Sci.}, vol. 18, pp 725-732, 1990;
M. Surendra and D. B. Graves, ''Particle simulation of
rf glow discharges,'' {\it IEEE Trans. Plasma
Sci.}, vol. 19, pp. 144-157, 1990.

\bibitem{6} G. L. Ragoff, J. M. Kramer, and R. B. Piejak, 
{\it IEEE Trans. Plasma
Sci.}, vol.14, pp. 103, 1986.
{\bf 30}, 641, 1959.

\bibitem{7} 
R. A. Gottscho and C. E. Gaebe,''Negative ion kinetics in RF
glow discharges,'' {\it IEEE Trans. Plasma Sci.}
vol. 14, pp. 92-102, 1986.
R. A. Gottscho and M. L. Mandich {\it
J. Vac. Sci. Tech Vol. A} vol. 3, pp. 617, 1985;
R. A. Gottscho, ``Glow-discharge sheath electric field:
Negative ion, power, and frequency effects,'' 
{\it Phys. Rev. A} vol. 
36, pp. 2233-2242, 1987.

\end {thebibliography}
\end{document}